\documentstyle[epsf]{article}

\newcommand{\muu}{\mbox{$\mu$}} \newcommand{\sig}{\mbox{$\sigma$}}
\newcommand{\muub}{\mbox{$\mu_0$}}
\newcommand{\sigb}{\mbox{$\sigma_0$}} \newcommand{\etal}{{\it et al.}}
\newenvironment{refs}{\section*{References} \begin{list}{}{
         \setlength{\parsep}{0pt} \setlength{\itemsep}{0pt}
         \setlength{\leftmargin}{0pt} } }{ \end{list} }

\title{Lognormal Properties of SGR 1806-20 and Implications for
  Other SGR Sources}
\author{K. J. Hurley, B. McBreen, M. Delaney and A. Britton}
\date{astro-ph/9508074: presented at 29 ESLAB Symposium, April 1995}

\begin{document}
\maketitle

\begin{abstract}
  The time interval between successive bursts from SGR 1806-20 and the
  intensity of these bursts are both consistent with lognormal
  distributions.  Monte Carlo simulations of lognormal burst models
  with a range of distribution parameters have been investigated. The
  main conclusions are that while most sources like SGR 1806-20 should
  be detected in a time interval of 25 years, sources with means about
  100 times longer have a probability of about 5\% of being detected
  in the same interval. A new breed of experiments that operate for
  long periods are required to search for sources with mean recurrence
  intervals much longer than SGR 1806-20.
\end{abstract}
\section{Introduction}

The lognormal properties of the soft repeater SGR1806-20 have been
previously reported by Hurley, K.J. \etal\ (1994). In particular, both
the time interval between repeater events and the luminosity function
of the source were fit with lognormal distributions (see Aitchison and
Brown, 1957, for a comprehensive introduction to lognormal
statistics). This analysis used the data-base of 111 events detected
by the International Cometary Explorer (ICE) mission, as reported by
Laros \etal\ (1987).

While the present number of events observed from the other two sources
(Norris \etal, 1991, Kouveliotou \etal, 1993) does not allow any
detailed analysis, the intervals between successive events of SGR
0526-66 (Golenetski\v{\i} \etal, 1987) is also suggestive of lognormal
behaviour. Continued observations by BATSE of these sources may reveal
lognormal properties for one or both of the remaining two repeaters if
either passes into a phase of activity similar to the behaviour of
SGR1806-20 during 1983.

The relationship between the number of active (i.e observable) sources
and the true number of SGRs in the galaxy is one which is the subject
of some debate (see discussions in Kouveliotou \etal, 1992,
Kouveliotou \etal, 1994 and Hurley, K. \etal, 1994). If the time
interval between SGR events proves to be lognormal then there may be
long quiescent periods where the source could be undetectable, leading
to an underestimate of the population.

\section{Simulations}

In order to investigate the behaviour of sources with much longer mean
recurrence times we generated Monte Carlo simulations with a variety
of distribution parameters. The Monte Carlo simulations were performed
using the random normal generator with Matlab 4.0 for Windows, which
is based on a random number generator algorithm given by Park and
Miller (1988) with the transformation to the standard normal variate
given by Forsythe, Malcolm and Moler (1977). The normal variates were
then transformed to lognormal variates using the relationship
$Y=e^{\sigma X+\mu}$ where $Y$ is lognormally distributed (with
parameters $\mu$\/ and \sig ) and $(\sigma \! X+\mu)$ is normally
distributed with mean $\mu$\/ and variance \sig.

\begin{figure}[thp]

  \epsfxsize=\textwidth \epsfbox{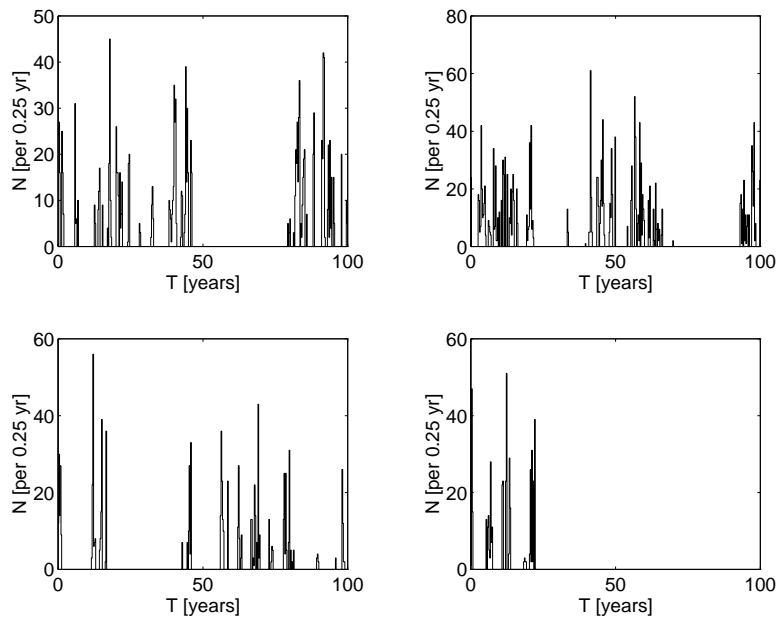}

  \caption{Four separate Monte Carlo simulations of SGR1806-20 activity over a
    hundred year period, generated from a lognormal distribution of
    recurrence intervals with the same parameters $\muub=9.64$ and
    $\sigb=3.44$. The long gaps in activity of the source are the
    contributions from the tail of this highly skewed distribution.}
  \label{sgr-sim}
\end{figure}

The parameters of the lognormal density function which were fit to the
distribution of recurrence intervals for SGR1806-20 were $\muub=9.64$,
$\sigb=3.44$ (Hurley, K.J. \etal, 1994). Initially we generated 100
year long simulations of SGR1806-20 (Fig.~\ref{sgr-sim}) using these
parameters, to check the algorithm.  The samples produced were tested
for compatibility with a lognormal population using a $\chi^2$ test
(Sachs 1986) and were compatible at the 99\% confidence level,
indicating that the Monte Carlo simulator was functioning correctly.
Two further simulations were then performed to investigate how the
source behaviour varied as \muu \/ and \sig \/ varied. The results
(illustrated in Fig.~\ref{probplot}) are discussed below.

\section{Discussion}

\begin{figure}[thp]
  \epsfxsize=\textwidth \epsfbox{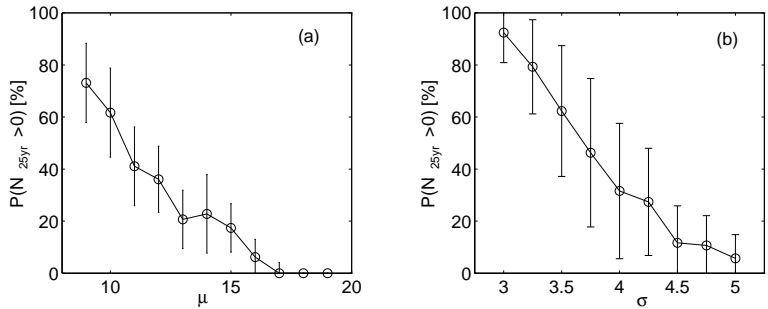}

  \caption{Percentage of time with one or more events per 25 year interval: (a)
    as a function of \muu ( with $\sigma = 3.0$ fixed) and (b) as a
    function of \sig (with $\mu=9.5$ fixed). Errorbars are 1 standard
    deviation values on the mean of 20 runs.}
  \label{probplot}
\end{figure}

Presented in Figure~\ref{probplot} are the probabilities for source
activity in a 25 year period as calculated from the results of the two
simulations described above. Figure~\ref{probplot}(a) shows that as
the parameter \muu increases the chance of one or more event in 25
years falls from $\approx 80\%$ at $\mu=9.5 \approx \mu_0$ (geometric
mean of 0.15 days) to less than 5\% at $\mu=18$ (geometric mean of
$\sim \! 500$ days).

The chance of one or more events in 25 years for a source like
SGR1806-20 (that is with $\mu\approx\mu_0 = 9.64$ and
\sig$\approx\sigma_0 = 3.44$) as predicted by Figure~\ref{probplot}
indicates that the majority of this type of source should be observed
in $\sim$ 25 years. For $\mu\!\gg\!\mu_0$ experiments which operate
for a long time must be devised and maintain a continuous search over
the whole sky for longer periods than any spaceborne experiments
designed so far. Such experiments could reveal a larger population of
sources with significant gaps of inactivity.

The lognormal distribution arises in statistical processes whose
completion depend on a product of probabilities, arising from a
combination of independent events (Montroll and Shlesinger, 1982).
Lognormal statistics have previously been used in connection with
gamma-ray bursts by McBreen \etal\ (1994) and Brock \etal\ (1994). In
the context of this investigation the physical significance of this
statistical behaviour may lie in the connection between SGRs and
neutron stars. In their paper, Hurley, K.J. \etal\ also presented a
similar statistical analysis of the behaviour of micro\-glitches from
the Vela pulsar (Cordes, Downes and Krause-Polstroff, 1988). The time
separation and the intensity of these small ( \(\mid \! \Delta\nu/\nu
\! \mid \sim 10^{-9}\) ) frequency adjustments were both compatible
with lognormal distributions, and there was no correlation between
waiting time and intensity: just as observed with SGR1806-20 (Laros
\etal, 1987). This result, combined with the identification of X-ray
point sources (Murakami \etal, 1994, Rothschild \etal, 1994) embedded
in plerion-powered SNR (Kulkarni \etal, 1993) as counterparts to the
SGR sources, suggests structural adjustments in neutron stars may be
the cause of SGRs.

\section{Conclusion}

Previously it was shown that the time intervals between successive events
from SGR1806-20 and the associated luminosity function were both lognormally
distributed. Structural adjustments in neutron stars may be responsible for
this behaviour. The activity of sources with longer mean recurrence times
was investigated using Monte Carlo simulations. The results of the
simulations indicate that there could exist a significant population of SGRs
with means longer than SGR1806-20 that remain undetected. A new breed of
experiments with very long observation times will be required to search for
this type of source.



\begin{refs}
\item Aitchison, J. and Brown, J.A.C., 1957, The Lognormal
    Distribution, Cambridge University Press: Cambridge.

\item Brock, M. \etal: 1994,
in Fishman, G.J.M, Brainerd, J.J, Hurley, K., ed(s), {\it A.I.P. Conf. Proc.
307,} 672.

\item Cordes, J.M., Downs, G.S. and Krause-Polstorff,
J., 1988, Astrophys. J., 330, 847.

\item Forsythe, G.E., Malcolm, M.A. and Moler C.B., 1977,
  Computer Methods for Mathematical Computations, Prentice-Hall.

\item Golenetski\v{\i}, S.V. \etal, 1987, Sov. Astron. Lett.,
  13(3), 166.

\item Hurley, K. \etal, 1994, Astrophys. J., 423, 709.

\item Hurley, K.J., McBreen, B., Rabbette, M. and Steel, S., 1994, Astron.
Astrophys., 288, L49.

\item Kouveliotou, C. \etal, 1992, Astrophys. J., 392, 179.

\item Kouveliotou, C. \etal, 1993, Nature, 362, 728.

\item Kouveliotou, C. \etal, 1994, Nature, 368, 125.

\item Kulkarni, S.R. \etal, 1994, Nature, 368, 129.

\item Laros, J.G. \etal, 1987, Astrophys. J., 320, L111.


\item McBreen, B., Hurley, K.J., Long, R. and Metcalfe,
L., 1994, MNRAS, 271, 662.

\item Montroll, E.W., Shlesinger, M.F., 1982, Proc Nat Acad Sci
USA, 79, 3380.

\item Murakami, T. \etal, 1994, Nature, 368, 127.

\item Norris, J.P., \etal, 1991, Astrophys. J., 366, 240.

\item Park, S.K. and Miller, K.W., 1988, Comm ACM, 32(10), 1192.

\item Rothschild, R.E., Kulkarni, S.R. and Lingenfelter, R.E., 1994, Nature
, 368, 432.

\item Sachs, L., 1986, Applied Statistics, Springer-Verlag: New York.,

\end{refs}
\end{document}